\documentclass[%
 reprint,
 amsmath,amssymb,
 prd,
]{revtex4-1}

\usepackage{graphicx}
\usepackage{dcolumn}
\usepackage{bm}

\begin{document}

\preprint{APS/123-QED}

\def\am{angular~momentum~}
\def\amm{angular~momenta~}
\def\al {\alpha}
\def\av{angular velocity~}
\def\avv{angular velocities~}
\def\as{asymptotically~}

\def\AAA{{\cal A}}  
 

\def\ba{\begin{eqnarray}}
\def\bam{\begin{array}}
\def\bA{\mbox{\bm$A\!\!$\ubm}~}
\def\bal{\mbox{\bm$\al$\ubm}~}
\def\bB{\mbox{\bm$B\!\!$\ubm}~}
 \def\bbb{background~}
\def\bbbb{backgrounds~}
\def\be{\begin{equation}}
\def\bj{\mbox{\bm$j\!\!$\ubm}~}
\def\bv{\mbox{\bm$v\!\!$\ubm}~}
\def\bet{\mbox{\bm$\eta\!\!$\ubm}~}
\def\bE{\mbox{\bm$E\!$\ubm}~}
\def\bof{\mbox{\bm$f\!\!$\ubm}~}
\def\bga{\mbox{\bm$\ga$\ubm}~}
\def\bi{\bibitem}
\def\bJ{\mbox{\bm$J$\ubm}~}
\def\bl{\mbox{\bm$l\!\!$\ubm}~}
\def\bm{\boldmath}
\def\bn{\mbox{\bm$n\!\!$\ubm}~}
\def\bna{\mbox{\bm$\nabla\!\!$\ubm}~}
\def\bOm{\mbox{\bm$\Om$\ubm}~}
\def\bR{\mbox{\bm$R$\ubm}~}
\def\bS{\mbox{\bm$S\!\!$\ubm}~} 
\def\bt {\beta}
\def\bT{\mbox{\bm$T$\ubm}~}
\def\br{{\bf r}}
\def\bz{\mbox{\bm$z$\ubm}~}
\def\bze{\mbox{\bm$\ze$\ubm}~}

\def\B {\overline}
\def\BB{\B B}
\def\BBB{{\cal B}}
\def\Bg{\Bar g}

\def\BK{\Bar {K}}
\def\Bna{\Bar\na} 
\def\BGa{\Bar\Ga}
\def\Br{\B r}

\def\ccc{cosmological~}
\def\cd{\!\cdot}
\def\cl{\centerline}
\def\co{coordinate~}
\def\coo{coordinates~}
\def\cy{cylinder~}
\def\cyy{cylinders~}

\def\CC{{\cal C}}

\def\de{\delta}

\def\dg{\dot{g}}
\def\dga{\dot\ga}
\def\di{\partial}
\def\dLL{\dot{\LL}}
\def\dna{ \dot{\na}}
\def\dom{\dot\om}
\def\dRR{\dot{\RR}}
\def\dW{\dot W}
 
\def\DD{{\cal D}}
\def\De{\Delta}

\def\ea{\end{eqnarray}} 
\def\ee{\end{equation}}
\def\eee{equation~}
\def\eeee{equations~}
\def\eM{electromagnetism~}
\def\ema{electromagnetic~}
\def\en{energy~}
\def\ep{\epsilon}
\def\eq{\equiv~}
\def\et{\eta}
\def \EE{{\cal E}}
\def\EEE{Einstein's~\eeee}
\def\ER{Einstein-Rosen~}

\def\fr{\frac}

\def\FF{{\cal F}}

 \def\ga{\gamma}
\def\gf{gravitational~field~}

 \def\gmn{g_{\mu\nu}}
\def\ggg{gravitational~}

\def\gm{gravomagnetism~}
\def\gmf{gravomagnetic~ $\!$field~}
\def\gmff{gravomagnetic~ $\!$fields~}
\def\gre{gravitational energy~}
\def\gw{gravitational wave~}
\def\gww{gravitational waves~}

\def\Ga{\Gamma}
\def\GG{{\cal G}}
\def\GR{General Relativity~}

\def\hmn{h_{\mu\nu}}
\def\ha{\frac{1}{2}~}
\def\hy{hypersurface~}
\def\hx{\hat{\xi}}
\def\HH{{\cal H}}


\def\iii{inertial frame~}
\def\iiii{inertial frames~}
\def\inf{\infty}

\def\k{\fr{1}{\ka}}
\def\kk{\fr{1}{2\ka}}

\def\II{{\cal I}}

\def\JJ{{\cal J}}

\def\ka{\kappa}
\def\kk{{\cal K}}
\def\kkk{Killing field~}
\def\kkkk{Killing fields~}
\def\kv{Killing vector~}

\def\KK{{\cal K}}

\def\la {\lambda}
\def\lb{\label}
\def\les{\ \lesssim}
\def\lhs{left-hand side~}
\def\lll{\left(}

\def\L{$\bullet\triangleright$~}
\def\La {\Lambda}
\def\LL{{\cal L}}
\def\LLL{\left[}

\def\mb{\mbox}
\def\me{mechanical energy~}
\def\mn{\mu\nu}

\def\Min{Minkowski~}


\def\na{\nabla}
\def\nl{\newline}
\def\nn{\nonumber}
\def\nnn{\noindent}
\def\np{\newpage}

\def\om{\omega}

\def\ov {\overline}
\def\Om {\Omega}

\def\OO{{\cal O}}
\def\pa{\parallel}
\def\po{\pounds}
\def\ppp{perturbation~}
\def\pppp{perturbations~}
\def\ps{\psi}

\def\PP{{\cal P} }

\def\QQ{{\cal Q }}

\def\ra{\rightarrow}
\def\rh{\rho}
\def\rhs{right-hand side~}
\def\rrr{\right)}
\def\rs{\rho\si}
\def\R{$~\triangleleft\bullet$~}
\def\RR{{\cal R}}
\def\Ra{\Rightarrow}
\def\RRR {\right]}

\def\s{{\it one$\!$}~}
\def\si{\sigma}
\def\sq{\sqrt}
\def\sqg{\sqrt{-g}}
\def\sqdg{\sqrt{--\ga}}
\def\sh{spherical harmonics~}
\def\sim{\simeq}
\def\sp{spheroid~}
\def\spp{spheroids~}
\def\ss{{\it two$\!$}~}
\def\sss{spacetime~}
\def\ssss{spacetimes~}
\def\st{stationary~}
\def\sup{superpotential~}
\def\sy{symmetric~}
\def\sym{symmetry~}

 \def\Si{\Sigma}
\def\Sc{Schwarzschild~}
\def\Sup{Superpotential~}
\def\SS{{\cal S}}
\def\td{\tilde}
\def\tDe{\tilde\De} 
\def\tEE{\tilde\EE}
\def\tga{\tilde\ga}
\def\tGa{\tilde\Ga}
\def\te{\theta}
\def\tf{\tilde f}

\def\ti{\times}
\def\tk{\td k}
\def\tLL{\tilde\LL}
\def\tm{\tilde m}
\def\tmu{\td\mu}
\def\tna{\tilde\na}
\def\tP{\tilde P}
\def\tQ{\tilde Q}
\def\tr{\td r}
\def\trh{\tilde\rho}
\def\tRR{\tilde\RR}
\def\ts{\textstyle}
\def\tsi{\td\si}
\def\tt{\tilde t}
\def\tte{\td\theta}
 \def\tV{\td V}
\def\tW{\td W}
\def\tw{\td w}

\def\Te{\Theta}
\def\TT{{\cal T}}
\def\TS{\thicksim}


\def\un{\underline}
\def\ubm{\unboldmath}
\def\und{\underbrace}

\def\UU{{\cal U}}


\def\vA{{\vec A}}
\def\vAAA{\vec{\AAA}}
\def\vB{\vec{B}}
\def\ve{\varepsilon} 
\def\vE{\vec{E}}
\def\vf{\varphi }
\def\vJ{{\vec J}}
\def\vom{\vec \omega}
\def\vOm{\vec \Omega}
\def\vp{\varpi}
\def\vr{{\vec r}}
\def\vna{\vec{\na}}
\def\vr{\varrho}
\def\vs{\vskip 0.5 cm}
\def\vt{\vartheta}
\def\vv{{\rm v}}
\def\V{{\rm V}}
\def\VV{{\cal V}}

   
\def\we{\wedge}
\def\wrt{with respect to~}

\def\WW{{\cal W}}

\def\YY{{\cal Y}}

\def\ze{\zeta}

\def\ZZ{{\cal Z}}

\def\1{{\it one}}

\def\smu{\cdot}

\def\2{{\ts{\ha}\!}}

\def\3 {\ts{\frac{1}{3}\!}}
\def\4{\ts{\fr{1}{4}\!}}
\def\6{\ts{\fr{1}{6}\!}}

\title{ Komar fluxes of  circularly polarized light beams and  cylindrical metrics}

\author{D. Lynden-Bell}%
 \email{dlb@ast.cam.ac.uk}
 \affiliation{Clare College, Cambridge \& Institute of Astronomy, University of Cambridge, Cambridge CB3 0HA, UK.}
\author{J. Bi\v{c}\'{a}k}%
 \email{jiri.bicak@mff.cuni.cz}
\affiliation{
Charles University, 180 00 Prague 8, Czech Republic}
%

\date{\today}
\begin{abstract}
 The mass per unit length of a cylindrical system can be found from its external metric
 as can its angular momentum. Can the fluxes of energy, momentum and angular momentum along the cylinder also be so found? We derive the metric of a beam of circularly polarized electromagnetic radiation from the Einstein-Maxwell equations.  We show how the uniform plane wave solutions miss the angular momentum carried by the wave. We study the energy, momentum, angular momentum and their fluxes along the cylinder both for this beam and in general. The  three Killing vectors of any stationary cylindrical system give three Komar flux vectors which in turn give six conserved fluxes. We elucidate Komar's mysterious factor 2 by evaluating Komar integrals for systems that have no trace to their stress tensors. The Tolman-Komar formula gives twice the energy  for such systems which also have twice the gravity. For other cylindrical systems their formula gives correct results. \\
 \end{abstract}

\pacs{Valid PACS appear here}
\maketitle


\section{Introduction}
Any system that carries a flux of angular momentum has an associated spatial twist in its metric. This was  illustrated earlier by a static cylindrical shell that carried a torque producing a flow of upward angular momentum upwards  which can equally be interpreted as a flow of downward angular momentum downwards \cite {LBK, LB}. From this  twisted metric  it was deduced that a flux of angular momentum along a cylindrical system could be detected via the external twist in its metric. However the externally twisted metric described above has $ d\vf dz$ and $dz^2$ terms that diverge at large radii.  Furthermore although there is a full discussion of the fecundity of Levi-Civita's metric which has three Killing vectors, the same arguments were not applied to the flat interior metric which was assumed to be $dt^2-[dR^2+R^2d\vf^2+dz^2]$. Thus the deduction that the angular momentum flux could be detected externally was not fully established. We shall return to this problem after we have gained understanding from a second cylindrical system that carries a flux of angular momentum,  the circularly polarized  beam of light.  The exact Einstein-Maxwell metric for the plane wave can be found in the literature see, for example, the fine paper \cite{vH}. However there is a problem with uniform waves that extend to infinite distance from their propagation axis. The electromagnetic vectors ${\bf E,~B}$ are perpendicular both to each other and to the propagation vector ${\bf k}$ so the Poynting vector $\boldsymbol{\Pi}={\bf E\ti B}/(4\pi c)$ lies along ${\bf k}$ so the angular momentum flux ${\bf R}\ti\boldsymbol{\Pi}$ has no component along the direction of propagation.  This is clearly wrong for circularly polarized waves. The apparent paradox is nicely resolved in Jackson's book \cite{Ja} Classical Electrodynamics examples 7.19 and 7.20 in the second edition or 7.28, 7.29 in the third, where it is shown that the angular momentum in the direction of propagation lies at the edge of the beam where the intensity falls off. It is also pointed out there that the ratio of the energy per unit length to the angular momentum per unit length of the beam is the angular frequency $\om$. 
This  agrees with the concept that each photon has an angular momentum $\hbar$ and an energy $\hbar \om $. (For the generalisation to linear gravity, see Barker \cite{Ba}.)

 To assess the angular momentum we must therefore consider non-uniform beams of finite cross section.
The metric of such a system was first considered by Bonnor in \cite{B1} and \cite{B2} but with the light replaced by null dust. Here we shall first detail the classical electromagnetic field of a circularly polarized beam in flat space. The angular momentum of a circularly polarized electromagnetic plane wave has been discussed for over a century. 

In \cite{Poy} Poynting considered the analogy between the mechanical model of a rotating shaft consisting of a thin cylindrical shell, and a beam of circularly polarized light falling normally on an absorbing surface, which led him to suggest that the beam would transfer its angular momentum to the surface. In a letter preserved in the Einstein archive at the Hebrew university of Jerusalem, Einstein wrote to the US National Research Council emphasising the importance of Beth's proposed experiment to demonstrate this. The experiment was a success; in~\cite{Be} Beth showed that a circularly polarized beam carried angular momentum just as Poynting had proposed.  An exact calculation of such a beam confined by a wave-guide was given in \cite{KK} while a general discussion of angular momentum in electromagnetic fields was given in \cite{St}. Discussions of the apparent paradox may be found in \cite{MC} where  the torque on a sphere embedded in such radiation is considered and more recently in \cite{BB} which relates the effects to quantum optics.

After detailing the stress-energy tensor of the electrodynamic beam we  derive the corresponding metric within general relativity. The solution is of the general form found in \cite{B2} for spinning null dust. We determine his free functions in terms of the electric field profile across the wave. For the beam of uniform intensity and finite cross section all the angular momentum is at the edge. Inside a rotating cylindrical shell space time is flat but in axes that rotate. We therefore conjectured that such a beam of circularly polarized radiation should give the same gravity field as the beam of unpolarized light given in Bonnor's  paper \cite{B1}, but in rotating axes.
We find this is so and thus demonstrate the relationship between Bonnor's papers.
Of course rotating axes have problems at large distances from the axis so although
they can be helpful locally, they are not part of a continuous global metric that can be
used externally far from the beam.
Later works \cite{FF}, \cite{Fr}  and  \cite{Po}  have concentrated on "gyratons", short bursts of waves as models of spinning particles. 

After giving the flat-space solution for a circularly polarized electromagnetic beam of finite cross-section in section 2, we discuss cylindrical boundary conditions for Einstein's equations in section 3. We give the general solution to the Einstein-Maxwell equations for such beams in section 4. In section 5 we give explicit solutions for beams with particular profiles. In section 6 we turn to the general questions posed earlier concerned with detection of conserved quantities and show how six conserved quantities may be detected asymptotically using the Komar integrals \cite{Ko}. We show that his formula, which agrees with Tolman's,  gives too large a result for the energy by a factor $2$ for beams of radiation while both give the correct result for static cylinders. We compare these results with others for cylindrical systems in the literature. In section 7 we revise the metric for the torqued cylindrical shell by a coordinate transformation that brings it into a form that obeys our boundary conditions but raises problems over the meaning of azimuthal angle $\vf$. We also show how a Komar  integral  allows us to use the asymptotic metric to calculate the flux of angular momentum carried by a torqued cylinder. 

For relativistic metrics we bring out the analogy between gravomagnetism and electromagnetism by using calligraphic letters $\AAA, \BBB, \EE, \HH,\DD$ for the gravitational fields that correspond to the ${\bf A, ~B,~E,~H,~D}$ of electromagnetism. These calligraphic symbols are vectors so should be considered as though they were in bold-face letters. We deal with axially symmetrical systems so we use a continuous azimuthal coordinate $\vf$ both inside the beam and outside it. We normally use the cylindrical coordinate $R$  defined so that $2\pi R$ is the circumference at fixed $z$ and $t$.

\section{Monochromatic circularly polarized light}
In flat space such a beam has  electromagnetic vectors that satisfy
\ba
{\bf\na\smu E}=0,~
\na^2{\bf E}=c^{-2}\di^2{\bf E}/\di t^2,~
{\bf\na\times E}=-\di{\bf B}/\di ct.
\ea
If the wave travels in the $z$ direction then ${\bf E}\propto \Re \exp[ik(z-ct)]$ and a solution of the equations (1) is
\ba
{\bf E}= \Re({\bf E_c}),~~~{\bf B}= -\Re(i{\bf E_c}),\nn\\
{\bf E_c}= E_0({\bf\hat{x}}+i{\bf\hat{y}}) \exp[ik(z-ct)],
\ea
where $E_0$ is a constant, a suffix c denotes a complex vector and unit vectors are denoted with hats. This electrodynamic field can 
also be described by the complex 4-vector potential $(A_{c})_t=0,~{\bf A_c= -E_c}i/k$. 
However such a wave fills all space and we wish to have a beam of finite cross section. So following Jackson \cite{Ja} we look  for a solution that falls to zero at the edge with $E_0=E_0(R),~~R^2=x^2+y^2$.
To satisfy ${\bf\na\smu  E}=0$ we take ${\bf E}=\Re\LLL[E_0({\bf\hat{x}}+i{\bf\hat{y}})+E_1{\bf\hat{z}}]\exp[ik(z-ct)]\RRR$ and find
\be
E_1=(i/k)(\di E_0/\di x+i\di E_0/\di y)=i(kR)^{-1}(x+i y)E_0',
\ee
where $E_0'=d E_0/d R.$
Thus our fields take the form
\ba
{\bf E}=\Re({\bf E_c}),~~~{\bf B}=-\Re[i{\bf E_c}],~~~{\bf A_c=-E_c} i/k,~~\nn\\
{\bf E_c}=[E_0({\bf\hat{x}}+i{\bf\hat{y}})+\frac{i\hat{{\bf z}}}{kR} E_0'(x+iy)] \exp[ik(z-ct)].
\ea
Because $E_0$ varies, (4) no longer satisfies (1) exactly, but provided $E_0(R)$ varies slowly so that it remains almost constant over the scale of one wavelength, then (4) remains an approximate solution with the amplitude varying slowly across the beam. The terms in equation (1) that we neglect are $(RE_0')'/R<< k^2E_0$ and the radial derivative of that inequality which is necessary for the ${\bf\hat{z}}$ terms. Thus provided $E_0$ varies only on the scale of the overall beam radius $R=a$ and $k a>>1$ the errors will be of order $(ka)^{-2}$.
Expressing the fields in real terms with $u=ct-z$,
\ba
 {\bf E} = 
 E_0[~{\bf\hat x}\cos(ku)+{\bf\hat y}\sin( -ku)] \nn \\+(E_0'/k) {\bf \hat z} \sin[-ku+\vf],~\\
 {\bf B}= E_0[-{\bf\hat x}\sin(ku)+{\bf\hat y}\cos(ku)] \nn\\+(E_0'/k){\bf \hat z} \cos[ku-\vf].~~
 \ea
 These fields make up the field tensor $F_{\mu\nu}=\di_\mu A_\nu-\di_\nu A_\mu$.
 How well do these approximate fields obey the conditions ${\bf E\smu B=0}$ and
 $E^2-B^2=0$ which ensure that the field's relativistic  invariants vanish?
 We find all the major terms vanish leaving only the squares of the edging fields
 which are of the order of those we neglect in the slowly varying approximation:
 \ba
 {\bf E\smu B}=-\2(E_0'^2/k^2)\sin 2[k(z-ct)+\vf],\\
 E^2-B^2=-\2(E_0^2/k^2)\cos 2[k(z-ct)+\vf];
 \ea
 both terms have $k^2$ in their denominators and average to zero over each half wavelength.
 The Poynting vector of the field (4) is 
\be
\boldsymbol{\Pi}=c{\bf E\ti B}/(4\pi)=\frac{c}{4\pi}\bigr[\frac{E_0E_0'}{kR}({\bf r\ti \hat{z}})+E_0^2{\bf \hat{z}}\bigl];
\ee
 the component around the beam has only one $k$ in its denominator. In keeping with the above we shall neglect terms of order $1/(ka)^2$ but keep terms of order $1/(ka)$. The $z$ component of angular momentum density $l_z$ is $ {\bf\hat{z}\smu (r\ti} \boldsymbol{\Pi})=({\bf\hat{z}\ti r})\smu\boldsymbol{\Pi}$, so
\be
l_z=-\frac{E_0E_0'R}{4\pi kc}.
\ee
For an almost uniform beam which falls off near the edge, $E_0'$ is zero except near the edge.
The total angular momentum per unit length along the beam  thus concentrated at the edge is
\be
L_z=-\frac{1}{4\pi kc}\int E_0E_0'2\pi R^2dR=\frac{1}{4\pi kc}\int E_0^2dV,
\ee
where we have integrated by parts from $R=0$ to the place where $E_0$ vanishes. The energy in unit length of beam is 
\be
U=\int\frac{E^2+B^2}{8\pi}dV=\int\frac{E_0^2}{4\pi}dV,
\ee
so $L_z=U/\om$ where $\om=kc$. This agrees with the quantum concept that each photon carries an angular momentum $\hbar$ and an energy $\hbar\om$.
The stress-energy tensor in Cartesian coordinates $(t,x,y,z)$ is given by\\\\
\begin{align}
4\pi \boldsymbol{T} = \begin{pmatrix}
 E_0^2 & -\Pi_x & -\Pi_y & -\Pi_z \\
 \cdot & 0      & 0      & -\alpha y \\
 \cdot & \cdot  & 0      & -\alpha x \\
 \cdot & \cdot  & \cdot  & -E_0^2 \end{pmatrix} \,,
\end{align}
  where $\al=E_0E'_0/(kR)$ and we have dropped the term $E_0'^2/k^2$ in the energy density
  as it is comparable with the terms neglected in our slowly varying approximation. The minus sign before the $\boldsymbol{\Pi}$ components arises because we use a + - - - signature and covariant spatial components of the tensor then have the opposite sign to contravariant ones which point in the direction of the vectors. For later use we put this tensor into cylindrical polar coordinates and use the relativist's coordinate components in the metric $ds^2=dt^2-(dR^2+R^2d\vf^2+dz^2)$:
\begin{align}
4\pi \boldsymbol{T} = \begin{pmatrix}
 E_0^2 & 0 & \alpha R^2 & E_0^2 \\
 \cdot & 0 & 0 & 0 \\
 \cdot&\cdot&0 &\alpha R^2 \\
 \cdot&\cdot&\cdot& -E_0^2 \end{pmatrix}
\label{}
\end{align}
  The other components are found by symmetry. Notice that the terms involving $\al$ are only significant close to the edge of an almost uniform beam. In the simplest model we treat them 
  as a delta function at the edge of a uniform beam giving surface currents
  \be
  J_0^\vf=\int T_0^\vf 2\pi RdR/(2\pi a)=-E_0^2/(4\pi k a)={\bf J\smu\hat z},
  \ee
  where ${\bf J}$  is the angular momentum per unit height in the beam. 
  We may summarise this energy momentum tensor as being the same as that for unpolarized light within the body of the beam but with a shell source at the edge that carries both a positive angular momentum and a torque. Hereafter we set $c=1$.
\section{Boundary conditions on cylindrical metrics}
For isolated systems boundary conditions are given via the notion of asymptotic flatness. This is in turn defined via conformal completeness at null infinity and appropriate initial data on a Cauchy surface at spatial infinity as introduced in the classical works of Penrose and others (see Wald \cite{Wa} for a review). Cylindrical systems are infinite by definition but one can exploit the asymptotic structure by identifying points along the $\di/\di z$ Killing vector and thus reducing the problem to a finite (2+1)-dimensional problem.This often  admits a conformal completion at null infinity and is flat near spatial infinity \cite{ABS};  however  in contrast to fully isolated systems, where stationary and radiative spaces satisfy the same boundary conditions,  the asymptotics for stationary cylinders is different from that of cylindrical waves  as explained in \cite{ABS} Appendix B;  for them standard conformal completion is not possible.\\
Alternatively following \cite{Wa} in 4-dimensions, we can define our boundary conditions for stationary cylindrical systems by the requirements that  there exist coordinates $(t,R,\vf,z)$ (in which $\vf$ is an azimuth and $R$ is the length of the corresponding Killing vector); further the metric components $g_{\vf \vf}=R^2$ by definition, $\xi^2=g_{t t}$ should be $O[R^n]$ for some $n,~~g_{t\vf}/\xi^2=O[1], ~~g_{t z}/\xi^2=O[1/R], ~~g_{R R}=O[R^n]$ for some $n,~~g_{\vf z}=O[R], ~~g_{z z}=O[\ln R]$ all at large $R$, wherever the system extends to large $R$. (Levi-Civita's metric with $m\ge 1$ does not.) We now adopt these boundary conditions generally.
  \section{Relativistic metric of circularly polarized light}
    An empty cylindrical shell of radius $a$ that carries torque has a flat static internal space but produces a $\vf, z$ twist ($g_{\vf z} \ne 0$) in its metric \cite{LB}.
  An empty rotating cylindrical shell produces a flat internal space but in axes that rotate
  relative to axes fixed at infinity.
  It is no surprise that an empty rotating cylindrical shell that carries a torque also produces
  a flat internal space in axes that rotate relative to infinity. We conjecture that the gravitational internal solution will be the same as \cite{B1} internal solution below. To agree with \cite{vH}  we write  $\Phi$ for Bonnor's $A$ which is  \emph{not} the electromagnetic vector potential considered above. For $R\le a$ we write
\be
ds^2=dt^2-(dR^2+R^2d\vf^2+dz^2)+\Phi(dt-dz)^2,
\ee
but the external solution will show that this is now relative to rotating axes. 
Equation (16) can be rewritten in the Landau and Lifshitz \cite{LL} form that completes the square on $dt$ and writes $\ga$ for the determinant of the spatial metric $||\ga_{kl}||$,
\ba
ds^2=\xi^2(dt-\AAA_k dx^k)^2-\ga_{kl}dx^k dx^l,~~~k,l=1,2,3\nn\\
~=\xi^2(dt-\Phi dz/\xi^2)^2-(dR^2+R^2d\vf^2+dz^2/\xi^2),\\
\Phi=\2\ka\si R^2,~~\xi^2=1+\Phi,\nn\\\AAA_z=\Phi/\xi^2,~~\ga=R^2/\xi^2,\nn
\ea
where $\si$ is the mass-energy density per unit cross sectional area of the beam.
The gravomagnetic induction in this metric is
\be
\BBB^k=(curl\AAA)^k=\ep^{klm}\di_l\AAA_m=[0,~-\ka\si/\xi^3,~0].
\ee
Here $\ep^{klm}$ is the alternating tensor that contains the factor $\ga^{-1/2}$. 
The usual gravitational acceleration is given by $\EE=-{\bf\na}\ln\xi$.  \cite{LBK} building on the work of \cite{LL} showed that in stationary spaces the gravomagnetic field, $\HH^k=\xi^3\BBB^k$, obeys an equation analogous to Maxwell's ${\bf curl~H}=4\pi{\bf j}$ to wit
\be
({\bf curl}\, \HH)^k=-2\ka\xi T_0^k,
\ee
so for the metric above, $\HH^k=(0,~-\ka\si,~0)$ which is purely toroidal as is appropriate for a current along ${\bf\hat{z}}$.  It is here treated as a 3-vector.  In relativity  $\HH^k$  is often called the twist vector. In terms of the Killing vector $\xi_\mu,~~~~\HH^\ka=-\ep^{\ka\la\mu\nu}\xi_\la\na_\mu\xi_\nu,~~~\HH^k$ consists of the spatial components of that 4-vector. Since $\xi_\ka\HH^\ka=0,~~~\HH^k$ determines the 4-vector. As we shall see later we are actually using axes that rotate at infinity so the Killing vector being used here is helical at large distances. With respect to the Killing vector that has no curl at infinity there is  a toroidal current too which leads to an extra $\HH$ field.\\
Bonnor's fine papers do not start from the Einstein-Maxwell equations.
Instead \cite{B1} treats unpolarized light as null dust and his \cite{B2} generalises this to some sort of spinning null fluid which he believed to have some relationship to a neutrino field. By contrast van Holten \cite{vH}   starts from the exact Einstein-Maxwell equations but unlike Bonnor he is interested in infinite waves that are of uniform amplitude perpendicularly to the wave vector rather than being confined to a beam. Since the edge is missing there is no angular momentum in these solutions; indeed they give the internal solution that we seek for a uniform beam, but need modification for application to a beam with a non-uniform amplitude $E_0(R)$. In \cite{B2}  he gives a metric which has sufficient generality to solve our problem. However he does not give an  interpretation of the physical meaning of his free functions or show how to specialise them  to be the metric of a circularly polarized light beam that satisfies the Einstein-Maxwell equations. We do this here. In fact Bonnor's results are considerably more general than those needed here.  His beams (and van Holten's) can be of finite length so they lack helical and $\di/\di z $ symmetry and although he is concerned with a beam that carries angular momentum, he does not specialise to axial symmetry and is therefore forced to consider angular momentum in the weak field approximation.  Beams of circularly polarized light do not have $\di/\di z$ or $\di/\di \vf$ symmetries in their electromagnetic fields which have only helical and null vector symmetries, however the Maxwell stress tensor of their electromagnetic fields have $\di/\di t,\di/\di\vf$ and $\di/\di z$ symmetries which therefore hold for the gravity of our circularly polarized beam. There are many  other examples of electromagnetic fields that do not inherit all the symmetries of the space-time they generate. These symmetries do not hold for the stress tensors of linearly or elliptically polarized beams and hold only approximately for unpolarized beams. \\
In what follows our $\Phi(R)$ (and van Holten's) is Bonnor's $A$, our $\psi$ is $\sqrt{2}$ times that of  Bonnor. Thus we write $\psi/\sqrt{2}$ where Bonnor writes  $\psi$. Also our $u$ and $\vv$ are $\sqrt{2}$ times those of Bonnor, so our $u=t-z,~~\vv=t+z,  ~~~du d\vv=dt^2-dz^2.$ Denoting the derivative of $\psi(R)$ by $\psi' $ his metric in our notation is
\be
ds^2=-dR^2-R^2d\vf^2+du d\vv+\Phi du^2-R\psi' d\vf du,
\ee
where the functions $\Phi,\psi$ depend on $R$ only. Comparing this to van Holten's exact Einstein-Maxwell wave which has the form (16) we conclude that $\psi$ is zero for the infinite uniform wave, so in the slowly varying approximation $\psi'$ is small and will be neglected when multiplied by $E_0'$. We also neglect $E_0''$ and $(E_0')^2$ as in the purely electromagnetic case of section 2. The metric (20) does not depend on $\vv$ which is an affine parameter along the null rays $(R,\vf,u)=$const. The null vector $l^\mu$ is Killing with only the component $l^\vv$ non-zero. It obeys $D_\al l_\bt=0$, so it is covariantly constant. The metric (20) belongs to the well-known class of $pp-$waves characterised by a non-twisting, non-expanding and non-shearing geodesic null congruence generated by the vector field $l^\mu$, see \cite{SKM}. We require a solution of the Einstein-Maxwell equations
\ba
D^\mu F_{\mu\nu}=0,~~\ep^{\mu\nu\ka\la}D_\nu F_{\ka\la}=0,\nn\\
G_{\mu\nu}=- \frac{\ka}{4\pi} [F_{\mu\la}F_\nu^{~\la}-\4~g_{\mu\nu}F^2].~~
\ea
The second equation in (21) is automatically satisfied by $F_{\mu\nu}=\di_\mu A_\nu-\di_\nu A_\mu$ and in $(t,x,y,z) $ coordinates we take ${\bf A}$ in the complex 4-vector form $(0,{\bf A_c})$
as given in (4). However we shall need this in $(R,\vf,u,\vv)=(x^1, x^2, x^3, x^4)$ coordinates in which it is $k^{-1}(-iE_0/k,E_0 R/k,-\2 E_0'/k^2,\2 E_0'/k^2)\exp(i\vf-i k u)$. With the metric (20) we find $ g=-\4 R^2$ and
\begin{align}
g^{\mu\nu}&=\begin{pmatrix}
 -1 & 0 & 0 & 0\\
 0 & -1/R^2 & 0 & -\psi'/R \\
 0 & 0 & 0 & 2 \\
 0 & -\psi'/R & 2 & \Phi^* \end{pmatrix} \\
\Phi^*&=-(4\Phi+\psi'^2). \nn
\end{align} 
From ${\bf A}$ we have the antisymmetric field tensor $F_{\mu\nu}$ where
\begin{multline}
F_{\mu\nu}\exp[-i(\vf+ku)]=\\
 \begin{pmatrix}
 0 & E_0'R/k & E_0 & 0 \\
   & 0       & iE_0R & 0 \\
   &         &  0& -\frac{1}{2} iE_0'/k \\
   & & & 0 \end{pmatrix}
\end{multline}
Using $g^{\mu\nu}$ to raise the indexes we find the contravariant components
\begin{multline}
F^{\mu\nu}\exp[-i(\vf+ku)]=
\begin{pmatrix}
0 & \frac{E_0'}{kR} & 0 & \frac{\psi'E_0'}{k}-2E_0 \\
  & 0               & 0 & -\frac{2iE_0}{R} \\
  &                 & 0 & \frac{2iE_0}{k} \\
  &                 &   & 0  
\end{pmatrix}
\end{multline}
Since  $F^{\mu\nu}$ is antisymmetric the first equation (21) becomes
\be
D_\mu F^{\mu\nu}=(2/R)\di_\mu(\2 R F^{\mu\nu})=0.
\ee
For $F$ given by (24), $E_0'$ is small and $kR$ is large so their ratio is neglected. The only surviving equation is that with $\nu=4$ which yields
\be
E_0'k^{-1}(\psi''+2\psi'/R)=0,
\ee
in which both terms are neglected in the slowly varying approximation and are exactly zero when the beam is uniform. Thus the Maxwell equations are satisfied in the curved space-time. The solution is exact for the uniform wave even when $\psi$ is present in the metric.
 We now turn to the Einstein equations. Since these are nonlinear in $F_{\mu\nu} $ we first put it
 into the real form using $\cos[\vf-k u]$ etc. We then find both $F^2=0=FF^*$ and
 \ba
 4\pi T_{\mu\nu}=F_{\mu\si}F^\si_{~~\nu}=  
\begin{pmatrix}
 0 & 0 & 0 & 0 \\
 \cdot & 0 & RE_0E_0'/k & 0 \\
 \cdot & \cdot & E_0^2 & 0\\
 \cdot & \cdot & \cdot & 0
\end{pmatrix}
 \ea
Bonnor gives the $T_{\mu\nu} $ corresponding to his metric (20) but with $x,y$ replacing $R,\vf $ and with the notational changes given earlier. Putting his result in our 
notation, using for instance $T_{u u}=\2[T_{u u}]_{Bonnor}$, we find his stress tensor has the form of equation (27) with
\ba
\ka T_{uu}=\2[\na^2\Phi+\4(\na^2\psi)^2]=2 G E_0^2;~~\nn\\
\ka T_{\vf u}=\ka T_{u \vf}=-\2 R\di_R\na^2\psi=2 G R E_0 E_0'/k.
\ea
This establishes that Bonnor's metric can be specialised to  correspond to the gravity of a circularly polarized electromagnetic beam in the high frequency limit. In what follows we treat the metric (20) with (27) as source exactly. We shall no longer throw away the small terms. Thus while the electrodynamics of a finite beam has to be done approximately (just as the solution of Maxwell's equations for a beam in flat space is approximate), the ensuing gravitational calculation will give the exact metric of the approximate stress tensor. The approximation becomes exact in both the high frequency limit  and for the uniform beam of infinite cross-section. We perform the integrations under the boundary condition that $R\psi'$ should not diverge at infinity. We then integrate the second equation of (28) to give
\ba
\na^2\psi=-2 G [E_0(R)]^2/k;~~~~~~~~~~~~~~~~~~\nn\\~\psi=-2 G \int_0^R \LLL \frac{1}{k R}\int_0^R E_0^2R d R\RRR d R.
\ea
Now $\2 \int_0^\inf E_0^2RdR$ is the total energy flux $S$ in the beam, ignoring its self-gravity, so the $\psi$ of equation (29) is only logarithmically divergent at large $R$. The first of (28)  gives
\be
\Phi=G\int_0^R \LLL\frac{1}{R}\int_0^R (4E_0^2-GE_0^4/k^2)~R d R\RRR dR,
\ee
which is logarithmically divergent too.
The formulae above give the metric for a circularly polarized beam of any profile. They satisfy the boundary conditions of section 3.
\subsection{An alternative metric  with a different boundary condition }
The metric above has  $\psi= -\2~G E_0^2 R^2/k$  for a uniform beam whereas van Holten's exact solution has it identically zero. If in place of the boundary condition used above we ask that $\na^2\psi=0$ on axis then from the second equation of (28) we see that
\be\na^2\psi=2G\De k,~~~~~~~~\De=[E_0^2(0)-E_0^2(R)].
\ee
Notice that $\De=0$ inside a homogeneous but finite beam.
 Integrating twice we find
\be
 \psi=2G\int_0^R[(kR)^{-1}\int_0^R\De R dR]dR.
 \ee
 Integrating the first of (28)
 \be
\Phi=4G\int_0^R \LLL\frac{1}{R}\int_0^R (E_0^2-\4 ~G\De^2/k^2)~R d R\RRR dR.
\ee
For reasons that will become clear we shall write this new metric with $\td{\vf}$
where we previously wrote  $\vf$ because they turn out to be different. The new forms of $\Phi$ and $\psi$  give an alternative metric for any circularly polarized beam. Outside the beam $E_0(R)$ drops to zero but  $\De=E_0^2 (0)$. Thus asymptotically $\psi\ra \2 ~G E_0^2(0)R^2/k$ and the term in the metric (20) becomes $-2\psi  d\td\vf du$ which has both a divergent $d\td\vf dt$ and a divergent twist term $d\td\vf dz$.
 If we set $\hat{\vf}=\td\vf+\Om t$ with $\Om=\psi/R^2=\2 G E_0^2(0)/k$, then the rotation at large $R$ would be eliminated. Thus the $\td\vf$ used in this alternative metric was actually an angle defined relative to axes that rotate at infinity and the $\hat{\vf}$ angles are the non-rotating ones. However that transformation would still leave a divergent twist term in the external metric. There is nevertheless an advantage in this alternative metric because for a uniform beam it gives $\De=0=\psi$ within the beam so it agrees with  van Holten's exact solution and shows that our conjecture at the start of section 3 has proved true. 
 Thus we have demonstrated that van Holten's solution when considered as the limit of a finite beam is actually in rotating axes and furthermore outside the finite beam the metric diverges like $R^2$ so it does not obey the boundary condition at large $R$.
 The Killing vector $\hat{\eta}=\di_{\hat{\vf}}$ obeys the regularity conditions at the axis so $\hat{\vf}$ can be identified with the azimuth everywhere not just at large $R$, but that would leave the very divergent twist term in the metric at large $R$ which is unacceptable. 
We now have two different metrics of the form (20), given by (29)+(30) and (32)+(33), describing almost the same physical situation, so we expect that they must be connected by a coordinate transformation. This is indeed true. If we write the first metric in terms of $\vf$ and then set 
\be
\vf=\td\vf+\Om u,
\ee
where $\Om= -\2GE_0(0)^2/k$, we get the   second form  of the metric. For uniform beams  this  transformation simplifies the  internal metric by eliminating the $R\psi'$ terms. However if we require that $\vf$ be continuous, as any true azimuth must be,  then it forces us into the unacceptably divergent form of the external metric. 
   These  difficulties of interpretation arise from the fecundity of a system with three different Killing vectors. The solutions of Einstein's local differential equations do not distinguish which combinations of $\vf, t, z$ are to be interpreted as the true angle about the axis, true time, and true coordinate along the axis. That interpretation comes from our imposition of the appropriate boundary conditions. 
In summary Van Holten's metric is perfectly good within the beam but it can not be extended to large $R$ outside a finite beam while keeping a continuous azimuthal angle. We shall show shortly how to amend it to avoid this problem.
\section{Light beams with specific profiles}
All these beams have metrics of Bonnor's form (20) or in cylindrical polars
\ba
ds^2=(1+\Phi)dt^2-[R\psi'dt d\vf+2\Phi dt dz+dR^2+\nn\\+R^2d\vf^2-R\psi'd\vf dz+(1-\Phi)dz^2],
\ea
where $\Phi$ and $\psi$ are functions of $R$, dependent on the beam's profile.
The metric can also be written in the Landau and Lifshitz \cite{LL} form
\ba
ds^2~=(1+\Phi)(dt-\AAA_\vf d\vf-\AAA_z dz)^2-\ga_{kl}dx^k dx^l,\nn\\~~
\AAA_\vf=\2 R\psi'/(1+\Phi),\AAA_z=\Phi/(1+\Phi),\nn\\~~\ga_{11}=1,~~
\ga_{22}=R^2[1+\Phi+\4\psi'^2]/(1+\Phi),~~~~~\\
~~\ga_{23}=-\2\frac{R\psi'}{1+\Phi},~~\ga_{33}=\frac{1}{1+\Phi},~~~\ga=\frac{R^2}{1+\Phi}.\nn
\ea
The usual gravitational acceleration is given by the 3-vector
\be
\EE_R=-\2\Phi'~{\bf\hat R}/(1+\Phi),
\ee
the gravomagnetic induction is given by the 3-vector
\be
\BBB^R=0,~~~\BBB^\vf=-\ga^{-1/2}\di_R \AAA_z,~~~\BBB^z =\ga^{-1/2}\di_R \AAA_\vf,
\ee
and the lines of gravomagnetic force by 
\be
\frac{\HH^\vf}{\HH^z}= \frac{\BBB^\vf}{\BBB^z}=-\frac{\di_R\AAA^z}{\di_R \AAA^\vf}=\frac{d\vf}{d z},
\ee
where we remember that $\HH=\xi^3\BBB$.
In what follows it is convenient to set the dimensionless
\be
s=(R/a)^2,~~~\mu=G[E_0(0)]^2a^2.
\ee
\begin{figure}[htbp]
\begin{center}
\includegraphics[width=7.5cm]{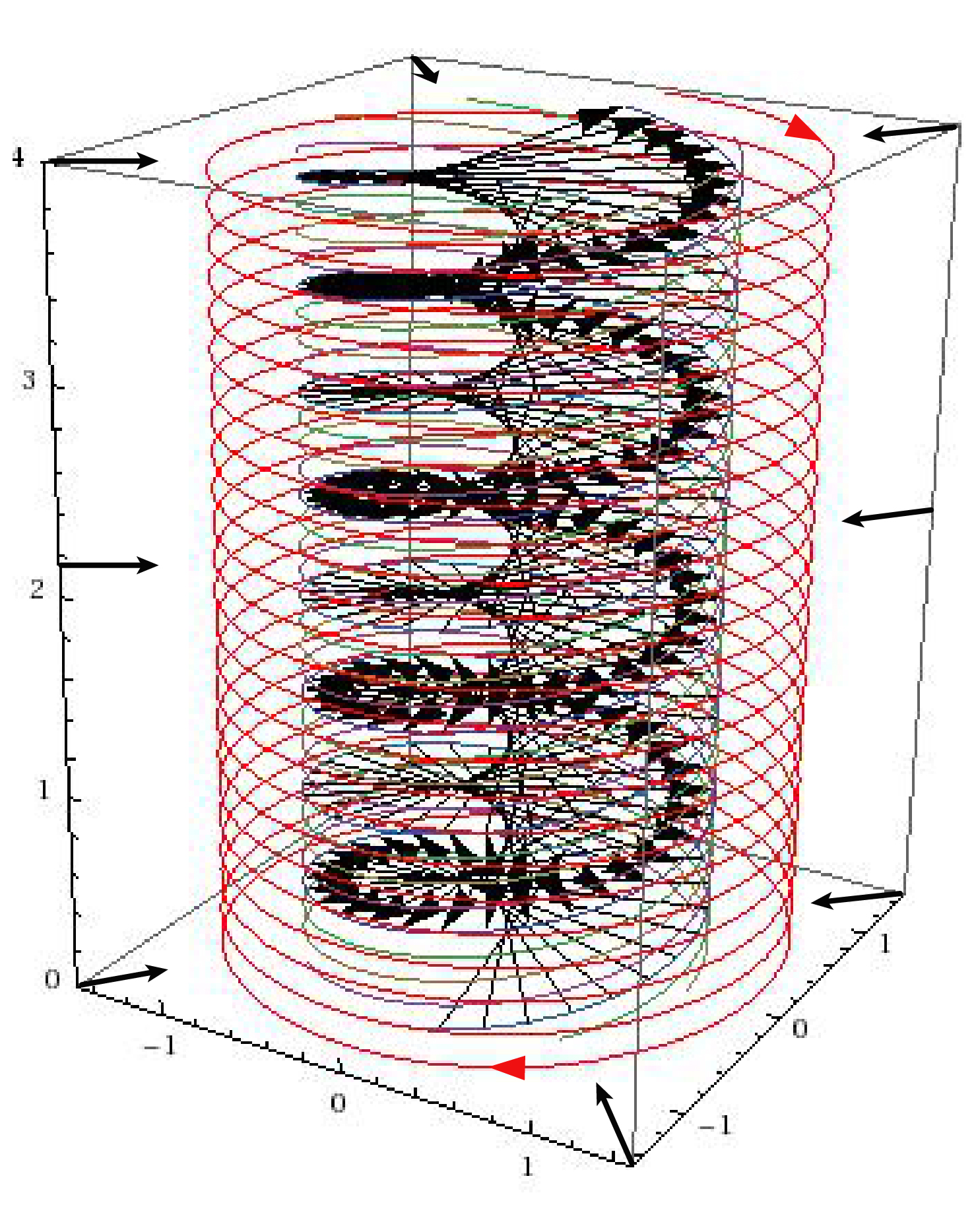}
\caption{The internal electric field of the circularly polarized beam of light is shown by the horizontal black arrows. The resulting external gravity by the inward radial  arrows and a gravomagnetic field line by the outer spiral.}
\label{fig.1}
\end{center}
\end{figure}
\subsection{The uniform beam of radius 'a'}
 The first form of the metric is given by (20) with $\psi$ and $\Phi$ given in (29) and (30). Then the uniform beam has
\ba
\psi=-\2 \mu s/k ,~~\Phi=\mu s[1-\4 \mu/(ka)^2] ,~~~R\le a;~~~\nn\\
\psi=-\2 (\mu/k)(1+\ln s),~~~~~~~~R\ge a;~~~~~~~~~~~~~~~\\
\Phi=\mu[1-\4 \mu/(ka)^2][1+\ln s]~~~~R\ge a.~~~~~~~~~~~~~~\nn
\ea
In this form of the metric $\Phi$ and $\psi$ only diverge logarithmically as does the Newtonian potential of a rod. Notice however that $\psi'$ is not zero outside the beam. Thus the twist caused by the transport of angular momentum along the beam is still detectable in the external metric. Setting $~Q=\mu[1-\4\mu/(ka)^2]$, the gravomagnetic field is
\ba
\HH^R=0,~~
\HH^\vf=-2\di_s\Phi/a^2=-2Q/a^2,~~R\le a;\nn\\ \HH^\vf=-2Q/(a^2 s),~~~~~~~~~~~~~~~R\ge a;\\
\HH^z=-\frac{\mu}{ka^2},~~R\le a; ~~~~\HH^z=\frac{\mu}{ka^2}Q/s,~~R\ge a.\nn
\ea
 The gravity field is radial and given by (37) with the new $\Phi$.
The gravomagnetic lines of force are helices 
\ba
d\vf/(k d z)=2Q/\mu=2[1-\4\mu/(ka)^2],\nn~~R< a;\\~~d\vf/(k d z)=-2/\mu,~~R> a.
\ea
These should be compared with the helix formed by the electric field with $d\vf/(kdz)=-1.$
Evidently the internal gravomagnetic field has its helix twisted in the opposite sense and with nearly twice the number of turns per unit height, while the external gravomagnetic helix is wound in the same sense as the electrical one but with many more turns per unit height. This external field is illustrated in Figure 1.\\
 In the second form of the metric
 \ba
 \psi=0,~\Phi=~\mu s ,~~R\le~a;\nn\\~~\psi= \2\frac{\mu} {k}(s-1-\ln s),~~R\ge a;\\
 \Phi= \mu[1+\ln s-\4\frac{\mu}{(ka)^2}(s-1-\ln s)],~~R\ge a.\nn
\ea
Notice that both $\Phi$ and  $\psi$ diverge like $s$ at large axial distance.
 Because we change to use the the time-like Killing vector of the internal space, the gravomagnetic field also changes to, in relativist's components, 
 \ba
\HH^\vf=-\Phi'/R=-2\mu/a^2,~~~\HH^R=\HH^z=0,~~~~R<a;\nn\\
\HH^R=0,~~\HH^\vf = -2\frac{\mu}{a^2}[1-\frac{\mu}{4(ka)^2}(s-1)],~R\ge a;~~~~~\\
\HH^z=\frac{\mu}{ka^2}\LLL1+\mu[1+(\ln s-1+1/s)(1+\frac{\mu}{4k^2a^2})]\RRR,\nn
\ea
with the purely radial gravity given by (37). The frame $\vf$ component is $R$ times the relativist's component and therefore diverges at infinity. The constant $\HH^\vf$ inside the beam corresponds to a physical component that grows like $R$. This is expected by Ampere's law for a uniform vertical current density. The vanishing of $\HH^z$ inside is related to the fact that inside a rotating cylindrical shell space-time is flat, but the inertial frame rotates relative to infinity. Inertial frames that have no rotation at infinity are more physical. In these we use a different Killing vector so the gravomagnetic field is not the same. In the form of the metric above both $\psi $ and $\Phi$ diverge as $O(R^2)$ at infinity. 
\subsection{Smoothly varying beams}
We choose $Ga^2 E_0^2(R)=\mu/[1+s^n]$. This is so constructed that it becomes the uniform beam as $n\ra\inf$, but varies smoothly for finite $n$.  We define for $0\le s\le 1$
\ba
F_1(s)=s-s^2/2^2+s^3/3^2-...\doteq s-\frac{\al_1s^2}{\bt_1+s}, \\
\bt_1=\frac{1-\pi^2/12}{1+\ln 2-\pi^2/6}-1=2.6822,\nn\\~\al_1=(\bt_1+1)(1-\pi^2/12)=.65371,\nn
\ea
and
\ba
F_2(s)=\int_0^s s^{-1}\ln(1+s) d s=F_1(s), ~ s\le1;\nn\\~
F_2=\pi^2/12+\2(\ln s)^2+\ln 2-F_1(1/s),~~s\ge 1.
\ea
Then for $n=1,~~\psi,~\Phi$ are given by
\begin{align}
\psi&=-\2(\mu/k)\int_0^s s^{-1}\ln(1+s) d s=-\2\mu F_2(s)/k,\nn\\
\Phi&=\mu F_2(s)-\4[\mu/(ka)]^2\ln(1+s).
\end{align}
For large $n$ we have likewise
\ba
\psi=-\2(\mu/k)s[1-n^{-2}F_1(s^n)],~~~~~~~~~~s\le1 ;~~~~~~~~~~~\\=-\2(\mu/k)\bigr[1-\frac{\pi^2}{6n^2}+(1+\frac{\pi^2}{6n^2})\ln s+\frac{s}{n^2}F_1(s^{-n})\bigl],\nn\\~s\ge 1;\nn
\ea
\ba
\Phi=\mu s\LLL 1-\frac{F_1(s^n)}{n^2}\RRR-~~~~~~~~~~~~~~~~~~~~~~~~~~~~~~~~~~~\nn\\-\frac{\mu^2}{4k^2a^2}s\LLL1-\frac{[s^n F_1(s^n)+\ln(1+s^n)]}{n^2}\RRR
~s\le 1 ;~~~~~\nn\\
=\mu\Bigl[1-\frac{\pi^2}{6n^2}+(1+\frac{\pi^2}{6n^2})\ln s+~~~~~~~~~~~~~~~~~~~~~\nn\\+\frac{s}{n^2}F_1(s^{-n})\Bigr]-\frac{\mu^2}{4k^2a^2}F_3,~~~~~~s\ge 1;~~~~\\
F_3= 1+\ln s+\frac{(\frac{\pi^2}{6}-\ln2)\ln s-\frac{\pi^2}{6}}{n^2}+~~~~~~~~~~~~~~\nn\\+\frac{s[F_1 (s^{-n})-\ln(1+s^{-n})]} {n^2}.~~~~~~\nn           
\ea
\section{Komar integrals for conserved quantities of stationary cylindrical systems}
Here we consider cylindrical systems which are invariant under the triple reversal
$t\ra -t,~z\ra-z,~\vf\ra-\vf$ so the metric has no cross terms involving  $dR$ and all the metric coefficients are functions of $R$ alone. There are then three Killing vectors
$\xi^\mu,\eta^\mu,\ze^\mu$ corresponding to shifts in the $t,\vf,z$ coordinates.  
From any vector field $V_\mu$ we can construct antisymmetric tensors and their symmetric counterparts
\ba
F_{\mu\nu}=\2(\di_\mu V_\nu-\di_\nu V_\mu)=\2(D_\mu V_\nu-D_\nu V_\mu),\nn\\
K_{\mu\nu}=\2(D_\mu V_\nu+D_\nu V_\mu).
\ea
Evidently
\be
D_\nu D_\mu V^\nu=D^\nu(K_{\mu\nu}+F_{\mu\nu})=D_\mu D_\nu V^\nu+R_{\mu\nu} V^\nu.
\ee
Thus
\be
D^\nu F_{\mu\nu}=R_{\mu\nu} V^\nu-D^\nu(K_{\mu\nu}-g_{\mu\nu}K);~~K=K^\mu_\mu.
\ee
For any Killing vector or much more generally  for any vector field that satisfies
$D^\nu(K_{\mu\nu}-g_{\mu\nu}K)=0$, we define a current with a minus sign only if $V^\mu$ is time-like. Inspection of the $T^\mu_\nu V^\nu$ term shows that this choice of sign
gives the fluxes in the right sense, hence
\be
\pm j^\mu=(1/\ka)D_\nu F^{\mu\nu}=(1/\ka)R^\mu_\nu V^\nu=-(T^\mu_\nu-\2\de^\mu_\nu T)V^\nu\,,
\ee
and  since $F_{\mu\nu}$ is antisymmetric, 
\be
\pm \ka D^\mu j_\mu=D^\mu D^\nu F_{\mu\nu}=(1/\sqrt{-g})\di_\mu\di_\nu(\sqrt{-g}F^{\mu\nu})=0.
\ee
Integrating over the 4-volume defined by $0\le t\le1,~~0\le z\le1,~~0\le\vf\le2\pi,~~0\le R\le\inf$, we get surface integrals of $j^\mu$ over the three dimensional faces of the 4-volume. Now provided $V_\mu$ satisfies the symmetries of the space (as it certainly will do if it is Killing) then the flux of $j^\mu$ through the two faces with $t=$constant will be equal and opposite as the normal points out of the 4-volume of integration. Thus the fluxes in the direction of increasing $t$ will be equal and give in $(t,R,\vf,z)$ coordinates
\ba
\int_0^1 \int_0^\inf\int_0^{2\pi}\pm  j^0\sqrt{-g}d\vf dR dz=\frac{2\pi}{\ka}\int_0^\inf\di_R(\sqrt{-g}F^{01})dR\nn\\=\frac{2\pi}{\ka}[\sqrt{-g}F^{01}]|^\inf.~~~~~~~~~
\ea
Likewise the flux through the two faces $z=$constant are equal and opposite and if evaluated in the direction of increasing $z$ give
\ba
\int_0^1 \int_0^\inf\int_0^{2\pi}\pm  j^3\sqrt{-g}d\vf dR dt=\frac{2\pi}{\ka}\int_0^\inf\di_R(\sqrt{-g}F^{31})dR\nn\\=\frac{2\pi}{\ka}[\sqrt{-g} F^{31}]|^\inf.~~~~~~~~
\ea
We shall apply this Komar technique to the currents $ j^\mu_{(\xi)},j^\mu_{(\eta)},j^\mu_{(\ze)}$, formed from each of the Killing vectors, $\xi^\mu,\eta^\mu, \ze^\mu$.
Notice that the currents $ \pm j^\mu=-(T^\mu_\nu-\2\de^\mu_\nu T)V^\nu$ are \emph{not} currents of the stress tensor itself but $T^\mu_\nu$ has twice its trace removed to make a trace-reversed stress tensor. Related to this, Komar finds that $ j^\mu_{(\xi)}$ must be multiplied by $2$ to give the mass, and his formula agrees with  Tolman's. This is a reflection of the fact that $T^0_0-\2(T^0_0 +T^k_k)=\2 (T^0_0-T^k_k), (k=1,2,3)$ and Komar's factor $2$ is needed to cancel out the resultant $\2$. No such factor is needed  for $j^\mu_{(\eta)}$  which yields the angular momentum because the trace term  $\2Tg_{\mu\nu}$ is not involved. We find below that Komar's formula for the mass per unit length of a light beam, which has no $T$, gives twice the mass, so his factor two, which was  compensating for the removal of half of $T^0_0$ by $-\2T$, is not needed when we deal with systems without a trace of $T^\mu_\nu$. Wherever
the Killing vector is in the direction of the surface over which the flux is evaluated  any non zero  trace $T$ affects the result. In other cases it does not because
the contribution to the integral from the trace is zero. For the special case of energy  Komar gives his extra factor $2$ which is the correct adjustment
for most stationary cases but, as we shall see, is wrong for a light beam which needs no such adjustment because $T=0$. However if the Tolman-Komar mass is not thought of as the energy content but as the gravitating power then it is always right since the beams with $T=0$
gravitate twice as strongly per unit energy as rods with there same $T_0^0$.
\subsection{Application to light beams.}
To define mass  or any other quantity per unit length  we need a definition of what we mean by unit length. For systems with a regular axis we introduce a coordinate time equal to the proper time there. In that spirit we introduce a local coordinate along the axis equal to the proper length there. We demand that the  axially symmetric coordinate $z$ be orthogonal to $t$ and $R$ everywhere. This allows us to extend $z$ away from the axis to the rest of space.
We may now talk of quantities measured per unit coordinate length.
To apply the Komar technique we need the inverse metric to (35) which is $g^{00}=1-\Phi-\4~ \psi'^2,~
g^{01}=0=g^{12}=g^{13},~g^{02}=-\2\psi'/R=g^{23},~~g^{03}=-\Phi,~~g^{22}=-1/R^2,~g^{33}=-(1+\Phi+\psi'^2), $ and the determinant $g=-R^2$.
From $j^\mu_{(\xi)}$ and equation (30) we obtain the mass per unit length, $M$, and the mass flux 
\begin{multline}
M=\frac{\pi}{\ka}[R\Phi']|^\inf=\\ 
\int_0^\inf \frac{[E_0(R)]^2}{4\pi}\LLL1-\frac{G [E_0(R)]^2}{4k^2}\RRR2\pi R dR.
\end{multline}
The $4\pi$ is due to the fact that the electric and magnetic fields have equal magnitude.
The term with $G$ is due to the self-gravitation of the electromagnetic beam.
Notice we have not used the formula advocated by Komar who uses $2j^\mu_{(\xi)}$
for the mass flux vector. His formula applied to this beam yields twice the mass per unit length. 
It is of interest that the gravity field flux vector $\DD_\mu=\xi\EE_\mu=\na \xi$, with $\xi=\sqrt{1+\Phi}$, so the gravitational flux toward the cylinder is $-\int\DD.dS=16\pi G M$ per unit length. Again twice the true mass per unit length. In this sense the energy in a light beam produces twice the gravity of the energy in a static cylinder.
The mass flux, $F_M$, is $c$ times (58) which is no surprise since the whole beam moves at that speed. Of course this is not the case for the rotating or the torqued cylinder considered later which have no mass flux along the cylinder.
From  $j^\mu_{(\eta)}$ we get the angular momentum per unit length
\be
J=-(\pi/\ka)[R\psi']|^\inf=k^{-1}\int_0^\inf(8\pi)^{-1}[E_0(R)]^2 2\pi R dR.
\ee
Again the angular momentum flux along the cylinder $F_J$ is just $cJ$. 
Finally from  $j^\mu_{(\ze)}$ we get the momentum per unit length, $P$, and its flux $F_P$,  the integral of the light pressure across the beam which  are hardly surprisingly $P=Mc$ and $Mc^2$.The fact that these expressions are so closely related to the energy is again due to the whole system moving at $c$ which ensures that $E=Pc$. 
A useful check on our result that the Komar formula with his factor two which is correct for static cylinders  gives twice the correct value for beams of light is provided by Bonnor's work on unpolarised light. Equation (5.1) of [B1] gives Bonnor's $A$ our $
\Phi=8M\ln(R/a)$ where $M$ is the mass per unit length; if we use the current $2j^\mu_{(\xi)}$ as advocated by Komar, we find Komar's formula gives $2\pi R \Phi'/\ka=2M$. Thus for these systems which have no trace $T$ we get the right mass using $j^\mu_{(\xi)}$ without the factor 2.
\subsection{Rotating waves and  rotating cylinders}
At the end of section 3 we concluded that van Holten's metric needed modification
if it was to be considered as the limit of a  metric of a laterally finite beam with a continuous azimuth. If we call his azimuth $\td\vf$, all we have to do is to apply the transformation (34) and rewrite his metric in terms of $\vf$ instead. This of course introduces a $d\vf dz$ term so the modified metric is of the form (20) with the new internal metric given by (41). While this looks more divergent at large $R$ than van Holten's original, this form only holds at $R\le a$ where the external forms  takes over. $R\psi' $ then converges so the only divergent terms in the external metric are those involving the logarithmically divergent $\Phi$. These obey our boundary conditions.\\ The metric outside a rotating cylinder of radius $b$ is
\ba
ds^2=\xi^2(dt^2-\AAA d\vf)^2-\xi^{-2}[e^{2k}(dR^2+dz^2)+R^2d\vf^2].\nn\\
\xi^2=[1-\om^2b^2(R/b)^{2m}],~~~~e^k=C(R/b)^{m^2},~~~~~~~~\\
\AAA=-\om b^2/(1-\om^2b^2)\nn \,.
\ea
See e.g. \cite{LBK} but note the sign change in the definition of $\AAA$ and  that the $R$ used here is not the length of the $\di/\di \vf $ Killing vector.
The contravariant metric reads $g^{t t}=\xi^{-2}-\xi^2\AAA^2/R^2,~g^{t \vf}=-\xi^2\AAA/R^2,~g^{\vf \vf}=-\xi^2/R^2,\\~~g^{R R}=g^{z z}=-\xi^2e^{-2k}$ and $\sqrt{-g}=RC^2(R/b)^{2m^2}\xi^{-2}$. From $j^\mu_{(\eta)}$ and the Komar expressions above we find the angular momentum per unit coordinate length of cylinder to be
\be
J=(2\pi/\ka)(-\AAA)(1-2m).
\ee
Precisely the same result follows from Bondi's definition of angular momentum in cylindrical systems \cite{Bo}. This definition is based on a quantity that remains conserved during slow changes of a cylindrical system and is apparently unrelated
to Komar's flux integrals. It therefore adds physical reality to Komar's definition. With no $d\vf dz$ terms in the metric $j^\mu_{(\eta)}$ has no component along $z$ so as expected there is no flux of angular momentum, $F_J$, along  cylinders in rotation
that carry no torque. 
Inside a rotating cylindrical shell of matter space-time is flat. However if the normal flat metric is fitted to an external one with a continuous $\vf$ then the external metric is in rotating axes and fails to obey our boundary conditions at large $R$. The external metric (60) obeys our boundary conditions; it fits the internal flat space metric when
the latter is written in rotating coordinates. The timelike Killing vector is not the same as in that of the non-rotating flat space. Indeed the new gravomagnetic field $\BBB$ is confined within the cylinder like the magnetic field of a long solenoid and it vanishes outside the cylinder since $\AAA$ in (60) is constant. However the flux through the cylinder is detectable externally through the interference of either electromagnetic or gravitational waves. As noted  by many authors, this gives a purely classical analogue of the quantum Bohm-Ahanarov effect in electromagnetism.
\subsection{Stressed cylindrical shells}  
A static cylinder of radius $b$ under longitudinal stress has no momentum but it carries a flux of linear momentum due to its stress. The flux vector of interest here is  $j^\mu_{(\ze)}$
and its flux through a constant $z$ surface. However we must again take care here because the stress (=momentum flux) that we wish to measure comes from $T_{\mu\nu}$ alone while the Komar integral will give us that plus a contribution from $-\2g_{\mu\nu}T$. The external metric can be put in the form  (2.7) in \cite{LBK}
 \ba
ds^2=\rho^{2nm}dt^2-[n^2C^2\rho^{2nm^2}dR^2+R^2d\vf^2+\rho^{-2m}dz^2],\nn\\~n=1/(1-m),~~\rho=R/b,~~~\xi=\rho^{nm}.~~~~~~~~~~
\ea
 Alternatively the metric may be taken in the form given by \cite{BLS} but notice that they use an $R$ that is not the length of the Killing vector $\eta$. Both give the same result.
The Komar flux per unit coordinate length is $F_\ze=2\pi m(1-m)/(\ka C)$. If we want the total stress we must add to this  $2\pi\int_0^\inf \2 T\ze\sqrt{-g}dR$. We cannot evaluate this using Komar integrals alone nor can we evaluate it using the asymptotic metric. We write the metric everywhere in the form
\be
ds^2=e^{-2\psi}dt^2-e^{2\psi}[e^{2(\ze-\al)}d\B R^2+\B R^2d\vf^2+e^{2\ze}dz^2],
\ee
where $\psi,\ze$ and $\al$ are zero at $R=0$ to ensure a regular axis with $z$ measuring proper distance along it. Notice that the metric outside the body where $\al$ is constant can be put into Weyl form only by changing $z$ by a constant factor, and $\psi=-m\ln(\B R/a), \ze=m^2\ln(\B R/b)$ there.
Adding the $G^{ R}_{R}$ and the $G^z_z$ components of Einstein's equations and multiplying by $\sqrt{-\B g}$, we find, writing a prime for $d/d\B R$
\be
(e^\al)'=\ka(p^R_R+p^z_z)\sqrt{-\B g}.
\ee
For shells $p^R_R=0$ so integrating and multiplying by $2\pi/\ka$ the momentum flux is
\be
F_P=2\pi(e^\al-1)/\ka.
\ee
This gives us directly what we want but the method only works when $p^R_R=0$ which is true for shells but is not generally true for static cylinders.
It is of interest to see how  stress affects the energy per unit length which is given via
 the flux of $2j^\mu_{(\xi)}$ through a constant $t$ surface. This gives
 $M=4\pi m/(\ka C)$ where the $m$ is that of Levi-Civita.
 In the classical limit ($C=1$) this gives the right result. It is also exactly the mass as given by the flux of gravitation $\DD=-\xi\na\ln\xi$. This gives a flux per unit length of cylinder of $-\int\DD.d{\bf S}=4\pi m/C$.
 
 \section{The torqued cylinder partially untwisted}
 The problem reconsidered here is a static cylindrical shell  that carries a torque. As explained earlier \cite{LBK,LB} such a system, though static, caries a flux of positive angular momentum upward which is the same thing as a flux of negative angular momentum downwards. It does this via a characteristically twisted external metric with a $d\vf dz$ term. The flat metric inside was taken to be the usual
 $ds^2=dt^2-[dR^2+R^2d\td\vf^2+dz^2]$, and defining $ l_1=\rho^{-2m}\cos^2\al+\rho^2\sin^2\al,~~~l_2=\rho^{-2m}\sin^2\al+\rho^2\cos^2\al,$ the external metric given in those papers is given by
 \ba
 ds^2-\rho^{2nm}dt^2=-[n^2C^2\rho^{2nm^2}dR^2+l_2b^2d\td\vf^2+~~~\\~~~~~~+b\sin2\al(\rho^{-2m}-\rho^2)d\td\vf dz+l_1dz^2]\nn\\
=-[n^2C^2\rho^{2nm^2}dR^2+\rho^2(h_1)^2+\rho^{-2m}(h_2)^2],\nn\\h_1=b\cos\al ~d\td\vf -\sin\al~dz,\nn\\ h_2=b\sin\al~d\td\vf+\cos\al~dz,\nn
 \ea
which is divergent at large $R$. However we can define a new azimuth $\vf=\td\vf-z\tan\al$ in terms of which the metric obeys the boundary conditions for cylindrical systems stated above. The external metric then becomes
 \ba
 ds^2=\rho^{2nm}dt^2-[n^2C^2\rho^{2nm^2}dR^2+R^2\cos^2\al~d\vf^2+~~\nn\\~~~~~+\rho^{-2m}(b \sin\al~d\vf+\sec\al~dz)^2].~~~~~~~~~
 \ea
 While this is fine and good the requirement that the azimuth be continuous with that of the internal metric forces us to make the same transformation there so the internal metric becomes
 \be
 ds^2=dt^2-[dR^2-R^2(d\vf+\tan\al~dz)^2+dz^2].
 \ee
 While we can see that this is a flat metric and check that  $\vf$ obeys the conditions
 required for an azimuth on axis nevertheless, it is disturbing to find that the surfaces of constant azimuth twist up around the $z$ axis at finite $R<b$ in the flat internal space.
 We are here forced to give up at least one of our ideas of what an azimuth should be:\\
 1) it should be continuous as we move in both $R$ and $z$.\\
 2) in a flat space the surfaces of constant $\vf$ and $z$ should be orthogonal.\\
 3) it should measure angle around the axis at both small and large $R$.\\
 To demonstrate that this form of static metric does indeed carry a flux of angular momentum we now calculate $-\pi \sqrt{-g}F^{31}_{(\eta)}$ from $j^\mu_{(\eta)}$
 in the metric (66).The contravariant metric is \\$g^{tt}=\rho^{-2nm},~g^{RR}=-n^{-2}C^{-2}\rho{-2nm^2},~g^{\vf\vf}=-1/(R^2\cos^2\al),~~~\\
 g^{\vf z}=-bR^{-2}\tan\al,~~~g^{zz}=-(\rho^{2m}\cos^2\al+R^{-2}b^2\sin^2\al)$ and $\sqrt{-g}=nCR\rho^{nm(1+m)-m}$. We deduce an angular momentum flux
 \be
 F_J=\2\pi(1-m^2)b\sin2\al~/(\ka C).
 \ee
\section{Conclusions}
i) We have expanded the use of Komar integrals for cylindrical systems and shown that a flux of angular momentum can be determined from the metric at large distances. This is also true of the energy flux but the momentum flux can be so determined only if $T=0$. Otherwise it can not be determined from the asymptotic metric.
We have cast some light on why Komar had to multiply the flux vector made from $\xi$ by a factor two to obtain the mass in agreement with Tolman's result \cite{To},  but no such factor was needed for the angular momentum flux made from $\eta$. If we think of the Tolman-Komar formula as giving an effective gravitating mass then we may regard it as always correct.\\
ii) Bonnor was right; his metric can be specialised to that for any circularly polarized beam of light.  We have determined his free functions in terms of the electric field and found the angular momentum  flux carried by the beam.\\
iii) The nice metric for the infinite beam found by van Holten is seen to be in rotating axes when it is considered as the limit of a finite beam, and we have modified it to obey
the boundary conditions of that application.\\
iv) We have amended the metric of the torqued static cylindrical shell by a coordinate transformation and demonstrated that it carries an angular momentum flux. However the amendment is not without some  damage to our concept of what is meant by an azimuthal angle, $\vf$. Perhaps even more challenging is the cylindrical NUT space discovered by Nouri-Zonoz \cite{NZ}.
\section{Acknowledgements}
 Stimulation for the study  of circularly polarized light came from the question of Prof Ue Li Pen after a seminar given at CITA in February 2016. \\
J.B. has support from the Czech Science foundation, Grant 14-370866 (Albert Einstein Centre) and is a Senior Visiting Fellow at the Institute of Astronomy, Cambridge. 
\\
\bibliography{polar-6}

\end{document}